\documentclass[12pt]{iopart}
\begin{document}
\def\beqn{\begin{eqnarray}}
\def\eeqn{\end{eqnarray}}

\title{A Note on Quantum Field Theories with a Minimal Length Scale}
 
\author{S.~Hossenfelder}

\address{Perimeter Institute, 31 Caroline St. N, Waterloo, Ontario, N2L 2Y5, Canada}
\ead{sabine@perimeterinstitute.ca}

\begin{abstract}
The aim of this note is to address the low energy limit of quantum field theories with a minimal length scale. 
The essential feature of these models is that the mininal length acts as a regulator in the
asymptotic high energy limit which is incorporated through an infinite series of higher order derivatives. If one investigates a perturbative expansion in inverse
powers of the Planck mass, one generically obtains extra poles in the propagator, 
and instabilities connected with the higher order derivative Lagrangian, that are however artifacts of 
truncating the series.  

\end{abstract}


\section{Introduction}	
 
The inclusion of gravity into quantum field theory appears to spoil renormalizability. It has therefore been suggested that gravity
should lead to an effective cutoff in the ultraviolet, i.e. to a fundamentally minimal length that provides
a natural regulator. It is amazing enough that all attempts towards quantum gravity indicate the 
existence  of such a minimal length scale, $L_{\rm m}$, which is expected to be close by, or
identical to the Planck length. Motivations for the occurrence of such a minimal length are manifold. It can be found in string theory, loop quantum gravity, non-commutative geometries, and it can be derived from a multitude of other approaches. 
For an overview, the interested reader is referred to \cite{Garay:1994en,Ng:2003jk,Hossenfelder:2004gj}.

Besides the attempts to pin down the emergence of a fundamentally finite resolution of spacetime,
the inclusion of a minimal length into the theoretical framework of the Standard Model (SM) has been examined as an effective description.  Such  models do not provide a full theory
of quantum gravity, but constitute a framework to study the phenomenological consequences of a minimal length scale. 
The aim of this note is to clarify the relation of this approach to SM extensions with higher order operators.

\section{Quantum Field Theory with a Minimal Length}

We introduce the existence of a fundamentally finite resolution by modifying the relation between wave-vector ${\bf k} = (\omega, k)$ and
momentum ${\bf p} = (E, p)$, in such a way that the wavelength can never become arbitrarily small, even at highest momenta. We thus introduce a non-linear relation ${\bf p}({\bf k})$ that has to fulfill the following requirements:

\begin{enumerate}
\item It is an odd function to preserve parity. \label{1}
\item The function is monotonically increasing, and invertible to ${\bf k}({\bf p})$. \label{2}
\item In the infrared limit it reduces to ${\bf p} = \hbar {\bf k}$. \label{3}
\item In the ultraviolet limit ${\bf k}$ becomes asymptotically constant at $1/L_{\rm m}$. \label{4}
\end{enumerate} 

Point (\ref{2}) implies that the function has exactly one zero, point (\ref{3}) means this zero is located at 
${\bf k}(0) = 0$. The most relevant feature of this model is point (\ref{4}), the existence of the asymptotic limit. This behavior can never be achieved with a finite order polynomial.  If one parameterizes the functions as ${\bf k} = (E f({\bf p}), p g({\bf p}))$ one can interpret this approach as an energy dependence of Planck's constant, and/or the speed of light defined by $\tilde \hbar({\bf p}) = 1/ f({\bf p})$ and $\tilde c({\bf p}) = c f({\bf p})/g({\bf p})$ \cite{Hossenfelder:2006rr}. Depending on the choice of $f$ and $g$ these models can, but need not necessarily have an energy dependent speed of light. The dispersion relation is modified to ${\bf p}({\bf k})^2 = m^2$, and upon quantization one obtains a generalized uncertainty principle (GUP) for $\Delta x \Delta p$. The function ${\bf p}({\bf k})$ is an input of the model, and different choices that fulfill the above requirements are possible. Frequently used versions are e.g. tanh or the Error function. The hope is that eventually a specific relation can be derived from a top down approach which would render this class of models a viable effective description.

It is understood here that the quantity ${\bf p}$ transforms as a standard Lorentz vector. It is then apparent that knowing the relation ${\bf k}({\bf p})$ results in a transformation for the wave-vector which, for a non-linear relation, will deviate from the standard Lorentz-transformation. In particular, this modified Lorentz-transformation needs to respect the asymptotic limit and the existence of the minimal length. It should be pointed out that these models do not break Lorentz invariance in the meaning that they do not single out a preferred reference frame. Instead they modify the Lorentz transformations at high boost parameters while preserving observer independence. These modifications of Special Relativity have become known as `Deformed Special Relativity' (DSR) \cite{Amelino-Camelia:2000mn,Magueijo:2001cr,Magueijo:2002am}. For more on the relation between these approaches, see \cite{Hossenfelder:2005ed}; for differences in the interpretation, see \cite{Hossenfelder:2006cw}. In some examples, point (\ref{4}) might hold only for either the spatial or the temporal component of the wave-vector. From hereon, we will assume that $f = g$ such that the speed of light remains a constant, and come back to the more general case in the discussion.  

The details of a quantum field theoretical model building up on this approach have been worked out in \cite{Hossenfelder:2006cw,Hossenfelder:2004up,Hossenfelder:2003jz}. Upon quantization, one obtains a higher-order derivative theory by essentially replacing the partial derivatives $\partial^\nu$ with higher order derivative operators defined through $\delta^\nu =  {\rm i} p^\nu( - {\rm i} \partial)$. E.g. for a scalar field $\phi$ the propagator becomes then the quantized version of the modified dispersion relation, and the Lagrangian reads ${\cal L}= (\delta^\nu \phi) g_{\nu\kappa} (\delta^\kappa \phi)$. 

As has been pointed out previously in \cite{Hossenfelder:2006cw} it follows from (\ref{4}) that the resulting field theory necessarily has an infinite order of derivatives. When integrating over a complete set of modes, the integration in ${\bf k}$-space has finite boundaries by construction. Transforming the integration into momentum space results in an additional factor stemming from the Jacobian $|\partial k/\partial p|$. This can alternatively be interpreted as working with a curved momentum space where this factor is the square root of the metric determinant. In the case of a constant speed of light, the propagator is multiplied by a function 
${\bf p}^2 = f({\bf k})^{-2}({\bf k}^2 - m^2)$, and momentum space is conformally flat. After Wick-rotation, integration over momentum space becomes finite since the minimal length acts as a regulator (which was the initial motivation for the model). 
As one might expect, models of this type contain an inherent non-locality at Planck-length level that reflects in the equal time commutation relations not identically vanishing outside the light-cone $[\pi(x), \psi(y)] \neq \delta(x-y)$, but being smoothened out over a Planck-length distance \cite{Hossenfelder:2007fy}. 

The relation between ${\bf k}$ and ${\bf p}$, which upon quantization yields the higher order derivative operator and the particle's propagator, uniquely specifies the model one is working with. The mentioned ambiguity in \cite{Magueijo:2006qd} for the choice of the Lagrangian arises only if one does not start from a fully defined {\sc DSR} (or, equivalently, with a relation ${\bf k}({\bf p})$), but instead with a modified dispersion relation. Since the latter is already a contraction, one has lost information that reflects in different possible Lagrangians which however represent different {\sc DSR} versions. This ambiguity is therefore only a consequence of not fully specifying the model. Further, the mentioned ambiguity in \cite{Bojowald:2004bb} for the Hamiltonian refers to the case in which Lorentz invariance is broken. As the authors state, Lorentz invariance uniquely relates the spatial with the temporal derivatives, and the ambiguity thus does not occur within {\sc DSR}. One way to deal with this is to define the Hamiltonian as the operator belonging to $E$, which will include higher order time derivates, and not as the operator belonging to $\omega$, as has been proposed in \cite{Hossenfelder:2003jz}. 

Theories with higher order derivative Lagrangians are known to suffer from instabilites, a missing lower-energy bound, and the initial value problem \cite{Jaen:1986iz,Pagani:1987ue,Eliezer:1988rr}. In these investigation however, the focus is on finite order derivative Lagrangians. In the model considered here, one sees that modifications of finite order could cause problems because the generic n-th order polynomial has n zeros. A finite number of derivatives would thus result in additional poles in the propagator. It would also make a larger amount of initial conditions necessary. The relation between both is apparent: the solution space to the wave-function is given by the zeros of the dispersion relation 
${\bf p}^2-m^2 = 0$, whose quantized version is just the propagator. 

In the here discussed case however, we are dealing with an infinite order power series, and the characteristic polynomial has exactly one zero on the real axis by construction.  Since the Lagrangian is hermitian, this means it is the only zero. This also implies no additional initial conditions are required. Physically seen this is apparent since the standard relation ${\bf p} = \hbar {\bf k}$ can be smoothly deformed into the nonlinear one, and the standard set of modes can be mapped to the modified one. There is no reason to expect new solutions to occur. For the more sophisticated version of these argument, see \cite{Barnaby:2007ve}.  

Writing down the expansion of the Lagrangian one obtains by replacing (gauge) derivatives with an infinite series with couplings in powers of $L_{\rm m}^2$, one arrives at a series of higher order operators, from which one could investigate the first some. However, with this one looses the essential feature that is the asymptotic high energy limit. Every single term in this expansion will contribute non-renormalizable higher order operators. This expansion is not very helpful for the same reason that a power series expansion to integrate over $\exp(-x^2)$ is not helpful, because it just results in an infinite series of divergent terms. As has been pointed out by Simon \cite{Simon:1990ic}, an infinite series expansion like the one considered here is subject to implicit constraints since the expansion is required to be convergent. Truncating the expansion at finite order would make it necessary to explicitly impose perturbative constraints on the resulting low-energy effective limit to stabilize the theory.

\section{Discussion}

In the more general case with an energy dependent speed of light, the analysis of the model's features becomes 
more involved because the modification of the propagator can not be written as a multiplication with a function. Thus, the pole structure needs to be examined separately for these cases. As mentioned previously, in some of the proposed {\sc DSR} versions, point (\ref{4}) does not hold for all components of ${\bf k}$. To understand the relevance of the minimal length in these cases, note that it is sufficient but not necessary for momentum space to have a finite volume to ensure 
loop contributions are non-divergent. The fact that deSitter-like momentum spaces are motivated by Loop Quantum Gravity, whereas other deformations emerge from particle scattering pictures with string-motivated modifications leads one to hope that the understanding of these models and their features would allow to pin down differences between both possible momentum space geometries.

It should be pointed out that the here mentioned models are so far not sufficiently understood, especially when it comes to the question of gauge invariance. A thorough investigation of the effective field theory limit is still missing, and the
relation to earlier technically similar, but conceptually different works \cite{Moffat:1990jj,Evens:1990wf} remains to be studied. However, we hope to have clarified that these points can in principle been addressed, hopefully in the soon future.

\section*{Acknowledgments}
 
I thank all participants of the workshop {\sl ``Experimental Search for Quantum Gravity''} held at Perimeter Institute, Nov 5th-9th 2007, for the interesting discussions.
Research at Perimeter Institute for Theoretical Physics is supported in
part by the Government of Canada through {\sc NSERC} and by the Province of Ontario through {\sc MRI.}

\section*{References}

{
 
}
\end{document}